\documentclass[twocolumn,prd,amsmath,amssymb,showpacs]{revtex4}
\usepackage{graphicx}
\usepackage{dcolumn}
\usepackage{bm}
\usepackage{epsfig}
\usepackage{amssymb}
\usepackage{amsmath}
\usepackage{color}

\def\fun#1#2{\lower3.6pt\vbox{\baselineskip0pt\lineskip.9pt
  \ialign{$\mathsurround=0pt#1\hfil##\hfil$\crcr#2\crcr\sim\crcr}}}

\renewcommand{\d}{\partial}

\newcommand{\bea}{\begin{eqnarray}}
\newcommand{\eea}{\end{eqnarray}}

\def\fun#1#2{\lower3.6pt\vbox{\baselineskip0pt\lineskip.9pt
\ialign{$\mathsurround=0pt#1\hfil##\hfil$\crcr#2\crcr\sim\crcr}}}
\def\lap{\mathrel{\mathpalette\fun <}}

\def\mpl{m_{\rm Pl}}
\newcommand{\MUNCH}[1]{\relax}

\begin{document}

\title{Direct Reconstruction of the Dark Energy Scalar-Field Potential}

\author{Chao~Li$^1$, Daniel E. Holz$^{2,3}$, Asantha~Cooray$^4$}
\affiliation{$^1$California Institute of Technology, Mail Code 130-33, Pasadena, CA~91125\\
$^2$Theoretical Division, Los Alamos National Laboratory,
Los Alamos, NM 87545, and\\
Department of Astronomy \& Astrophysics,
University of Chicago, Chicago, IL 60637\\
$^3$The Observatories of the Carnegie Institution of
Washington, Pasadena, CA 91101\\
$^4$Center for Cosmology, Department of Physics and Astronomy, University of California, Irvine, CA~92697}

\begin{abstract}
While the accelerated expansion of the Universe is by now well
established, an underlying scalar field potential possibly
responsible for this acceleration remains unconstrained.  We
present an attempt to reconstruct this potential using recent SN
data, under the assumption that the acceleration is driven by a
single scalar field. Current approaches to such reconstructions
are based upon simple parametric descriptions of either the
luminosity distance or the dark energy equation of state (EOS). We
find that these various approximations lead to a range of derived
evolutionary histories of the dark energy equation of state
(although there is considerable overlap between the different
potential shapes allowed by the data). Instead of these indirect
reconstruction schemes, we discuss a technique to determine the
potential directly from the data by expressing it in terms of a
binned scalar field. We apply this technique to a recent SN
dataset, and compare the results with model-dependent approaches.
In a similar fashion to direct estimates of the dark energy
equation of state, we advocate direct reconstruction of the scalar
field potential as a way to minimize prior assumptions on the
shape, and thus minimize the introduction of bias in the derived
potential.
\end{abstract}
\pacs{98.80.Es, 97.60.Bw, 98.80.Cq}

\maketitle

\section{Introduction}
Distance estimates to Type Ia supernovae (SNe) are currently a preferred probe of the expansion
history of the Universe \cite{Rieetal04}, and have led to
the discovery that the expansion is accelerating
\cite{perlmutter-1999}. It is now believed that a  mysterious dark energy component,
with an energy density $\sim$70\% of the total
energy density of the universe, is responsible for the accelerated expansion \cite{Spergel}.
While the presence of acceleration is now well established by various cosmological probes,
the underlying physics remains a complete mystery. As the precise nature of the dark
energy has profound implications, understanding its properties is one of the
biggest challenges today.

With the advent of large surveys for Type Ia supernovae,
such as the Supernova Legacy
Survey (SNLS)~\footnote{http://www.cfht.hawaii.edu/SNLS/} and
Essence~\footnote{http://www.ctio.noao.edu/~wsne/}, among others,  it is
hoped that we will study details of the expansion, and
thereby elucidate the
physics responsible for the acceleration. Under the assumption that the
dark energy is due to a single scalar field rolling down a potential,
several studies have considered how future data might be used to
reconstruct the potential, either based on various
analytical descriptions of the luminosity distance
\cite{HT99}, or through specific
assumptions about the potential, such as a polynomial function in the scalar field \cite{Sahlen}.
It is already well established that certain parametric descriptions of the distance
lead to biased estimates for the dark energy equation-of-state (EOS) and the potential \cite{Weller}.
While improved parametric forms of fitting functions have
been suggested \cite{Gerke,Guo}, it is unclear how to select
an optimal approach for reconstructing the dark energy
scalar field potential from SN distances (for a review
of various possibilities, see Ref.~\cite{Sahni:06}).

In this paper we discuss issues related to potential and dark
energy EOS reconstruction by making use of a recent set of SN data
from the SNLS survey \cite{Astier}. The sample includes 73 high
redshift SNe complemented with a sample of 44 nearby supernovae
\cite{Astier}. We compare and contrast a variety of methods to
reconstruct the potential and the dark energy EOS. We write the
luminosity distance either as a simple polynomial expansion in
redshift, or as a Pad\'e approximation \cite{Saini} (which avoids
some of the known problems in the polynomial expansion when taking
derivatives \cite{Huterer2,Weller,Gerke}). In addition to
approximating the luminosity distance, we also explore two
approximations to the EOS: $w(z)=w_0+w_a(1-a)$
\cite{David,Linder03} and $w(z)=w_0-\alpha \ln(1+z)$ \cite{Gerke}.

Based on our model reconstruction of the potential,  we find  that
while there is significant overlap of the allowed $V(\phi)$ region
favored by each of the four reconstruction methods, the models
give rise to different histories for the EOS, especially within
the two parameter plane, $w$--$w'$ (the EOS parameter, $w$, and
its time derivative, $w'\equiv dw/d\ln a$, as functions of
redshift \cite{Caldwell}).  We argue that existing parametric
fitting functions for either distance or the EOS lead to biased
reconstructions of the potential. In the literature, however,
there exist model-independent approaches to the reconstruction of
the dark energy density \cite{Wang} and the EOS \cite{HutCoo},
which bin the parameters directly as a function of redshift, with
the number and width of the bins determined by the statistical
quality of data. These estimates can also be arranged to be
uncorrelated \cite{HutCoo}, allowing unique insights into the
evolution without being subject to prior assumed redshift
dependencies. Here we suggest a similar model-independent approach
to the reconstruction of the scalar potential from SN data.
Instead of utilizing a polynomial expansion for the potential
\cite{Sahlen}, which assumes a limited range of models (once the
expansion is truncated at a certain order),  we propose a binning
scheme for the potential that can be applied to data with a
minimal, and easily controlled and understood, number of
assumptions for the potential shape.

The paper is organized as follows: In the next Section we review
techniques for reconstructing the scalar-field potential from SN
distances. We also reconstruct the EOS as a function of the
redshift, and use this to study the $w$--$w'$ plane (which has
been advocated as a way to characterize the underlying potential
responsible for the dark energy component by separating the regime
into ``freezing'' and ``thawing'' potentials \cite{Caldwell}; see,
also \cite{HutPei} for a Monte-Carlo exploration). In Section~III
we explore the impact of different parameterizations on the
derived evolutionary histories. While we observe these differences
with $\sim$ 115 SN data points, future large SN datasets may lead
to apparently inconsistent results. In Section~IV, following the
approach to model-free estimates of the dark energy EOS
\cite{HutCoo},  we present a model-independent estimate of the
scalar field potential. We conclude with a summary of our main
results in Section~V.

\section{Potential Via Parametric Forms}
\label{parametric}

For this study we make use of SN data from SNLS \cite{Astier}. Due
to complications related to independent data sets (e.g., differing
calibration, color correction, extinction correction, etc.), we do
not attempt to increase the sample size by combining other SN
datasets.  The measurements from Ref.~\cite{Astier} present the
quantity $\mu_B=m_B-M$ for 117 SNe, with 73  of these at redshifts
greater than 0.2 \footnote{The distance estimates for two
high-redshift SNe lie more than $3\sigma$ away from the best-fit
relation in the Hubble diagram. As in Ref.~\cite{Astier}, we
exclude these two data points and only make use of 115 data points
to model fit the data.} This distance modulus is related to the
luminosity distance through $\mu_B=5\log_{10}d_L$, while the
luminosity distance is related to the comoving radial distance via
$d_L=c(1+z)r(z)/H_0$, where $r(z)=\int_0^z dz'/H(z')$ with $H(z)$
the expansion rate of the Universe. When model fitting the data,
we fix ${M}=19.3 \pm 0.03$  to the value determined by SNLS. We
take the central value; further uncertainty will be incorporated
into $\sigma_{\rm int}$, as discussed below.

\begin{figure}[tb]
\includegraphics[scale=0.33,angle=-90]{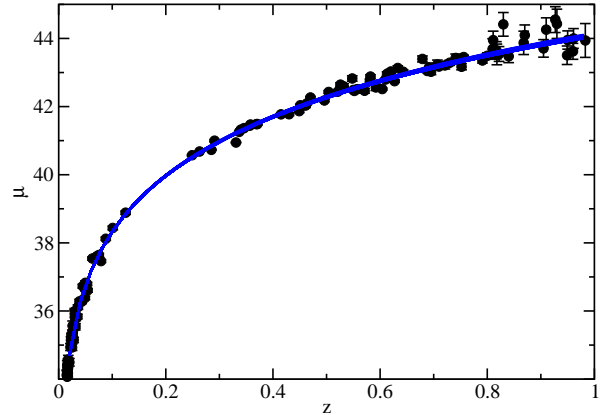}
\caption{Hubble diagram for 115 Type Ia SNe used in the present analysis. The error bars are
on $\mu_B$ only. We also include an additional constant
error, $\sigma_{\rm int}=0.13$, to account for the SN intrinsic dispersion. For reference
we also plot 300 curves drawn uniformly from the 2$\sigma$
consistent likelihood
fits to the data using the Taylor expansion with $r(z)=z+a_2z^2+a_3z^3+a_4z^4$.
}
\label{fig:one}
\end{figure}

In our reconstruction of the potential, we describe $r(z)$ through
two parametric forms widely used in the literature. First, we
expand $r(z)$ as a simple power-law \cite{HT99} such that
\begin{equation}
r(z)=z+a_2z^2+a_3z^3+a_4z^4  \, .
\end{equation}
Note that the coefficient of the first order term is exactly one.
Since this polynomial expansion has known problems when estimating
the derivatives of $r(z)$ (e.g., Figure~3 of Ref.~\cite{Gerke},
and also Ref.~\cite{Jonsson}), we also consider a Pad\'e form for
$r(z)$ with  Ref.~\cite{Saini}:
\begin{equation}
r(z) = 2 \frac{z+c_1(1-\sqrt{1+z})}{c_2(1+z)+c_3\sqrt{1+z}+2-c_1-c_2-c_3} \, ,
\end{equation}
such that as $z \rightarrow 0$, $r(z) \rightarrow z$.
In this form, using $r(z \rightarrow \infty)$, one can
additionally constrain the parameters with:
\bea
3 \Omega_M&\le &\frac{4c_2+2c_3-c_1}{2-c_1}\nonumber\\
1\le&{1\over c_2}&\le
\frac{1}{2}\int_1^\infty\frac{dx}{\sqrt{1-\Omega_M+\Omega_M x^3}}
\, . \eea In addition to the two fitting forms for $r(z)$, we also
determine $r(z)$ through model parameterizations for $w(z)$,
including $w(z)=w_0+(1-a)w_a$ \cite{David,Linder03} and
$w(z)=w_0+\alpha \ln (1+z)$ \cite{Gerke}. Since from $w$ it is
possible to determine the distance, these approximations allow us
to once again reconstruct the dark energy potential.

In each of two parametric descriptions of $r(z)$ we have three free parameters.
We parameterize $w(z)$ with two parameters, and include $\Omega_m$ as a third
free parameter (under the assumption of
a flat universe; weakening this assumption significantly degrades our ability to
measure anything about the potential with existing data). When showing results
related to potentials or EOS as a function
of redshift, we take a prior on $\Omega_m$ such that the
probability is Gaussian with a mean of 0.25 and a standard deviation given by
$\sigma=0.05$ \cite{Spergel}.
In each case, to obtain the join likelihood distribution of the parameters given the data,
we perform a likelihood analysis:
\begin{equation}
\chi^2(p_i)=\sum_{i=1}^{N}\frac{[\mu-\mu_{B}(z_i)]^2}{\sigma_{\mu_B}^2+\sigma_{\rm int}^2} \, ,
\end{equation}
where, following Ref.~\cite{Astier}, in addition to statistical
uncertainty in $\mu_B$ we include an additional Gaussian
uncertainty, $\sigma_{\rm int}=0.13$, representing the intrinsic
dispersion of SN absolute magnitudes, $M$.  We ignore
complications related to covariances in the Hubble diagram, either
due to effects related to calibration  \cite{kim} or fundamental
limitations such as gravitational lensing correlation of SN flux
\cite{Cooray} or peculiar velocities \cite{Hui}. The posterior
probability distribution is taken to be $P(p_i|\mu)\propto
e^{-\frac{1}{2}\chi^2(p_i)}$, and we marginalize the likelihood
over the uncertainty in $\Omega_m$, assuming a Gaussian prior
distribution.

Once the joint probability distribution for parameters is
determined, we sample the 1$\sigma$ and 2$\sigma$ range allowed by
these parameters to draw a fixed ($> 600$) number of independent
$r(z)$ curves consistent with the data.  For each of these
distance curves,  $r_i(z)$, we obtain the scalar-field potential,
in dimensionless units such that
$\tilde{V}(\tilde{\phi})=V(\phi)/\rho_{\rm crit}=V/(3H_0^2/8\pi
G)$,  through Ref.~\cite{HT99} \bea
\tilde{V}(\tilde{\phi})&=&\left[\frac{1}{(d\tilde{r}/dz)^2}+\frac{1+z}{3}\frac{d^2\tilde{r}/dz^2}
{(d\tilde{r}/dz)^3}\right] -\frac{1}{2}\Omega_M(1+z)^3 \, , \nonumber \\
\eea where $\tilde{r}=H_0r$. For each of the $r_i(z)$ estimates,
we also randomly draw $\Omega_m$ from a Gaussian prior
distribution as described above.  The mapping between $z$ and
$\phi$,  the scalar field value, is obtained through \bea
\frac{d\tilde{\phi}}{dz}&=&-\frac{d\tilde{r}/dz}{(1+z)}\nonumber\\
&\times&\left[-\frac{1}{4\pi}\frac{(1+z)d^2\tilde{r}/dz^2}{(d\tilde{r}/dz)
^3}-\frac{3}{8\pi}\Omega_M(1+z)^3\right]^{1/2} \, ,\eea where
$\tilde{\phi}=\phi/\mpl$. Furthermore, for models where we
parameterize $r(z)$, we can also extract the dark energy EOS as
\bea
w(z)=\frac{1+z}{3}\frac{3\Omega_m(1+z)^2+2(d^2r_i/dz^2)/(dr_i/dz)^3
}{\Omega_m(1+z)^3-(dr_i/dz)^{-2}}-1 \,. \nonumber\\
\eea When selecting models associated with scalar fields, we
require that $d \tilde{\phi}/dz >0$, such that $w \geq -1$. Even
in the case of $w(z)$ parameterizations where model fits allow $w
< -1$, we ignore $w(z)$ below this value as single scalar-field
models do not naturally give rise to such EOS.

\begin{figure}[tb]
\includegraphics[scale=0.4,angle=-90]{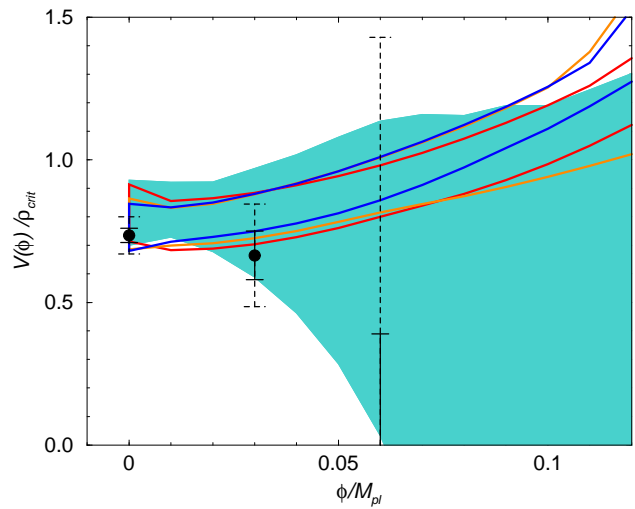}
\caption{The normalized quintessence potential $\tilde V(\phi)$ vs $\phi/\mpl$.
The shaded region is allowed at the 2$\sigma$ confidence level when using the
Taylor expansion for $r(z)$. The red lines mark the same when using a Pad\'e approximation to
the distance. The orange and blue lines are for the cases where $w(z)$ is
parameterized by
$w(z)=w_0+\alpha \ln(1+z)$ and $w(z)=w_0+w_a(1-a)$, respectively. Here $\phi=0$ corresponds to
$z=0$, while $\phi > 0.1$ generally corresponds to $z > 1$ (depending on $d\phi/dz$). The points with error
bars show the $1\sigma$ (solid) and $2\sigma$ (dashed) model-independent
estimates of the potential described in Section~IV (see equation~(8)).
While there is considerable overlap in the allowed region,
there are also significant differences in terms of the redshift evolution of
the EOS.}
\label{fig:two}
\end{figure}

\begin{figure}[tb]
\includegraphics[scale=0.4,angle=-90]{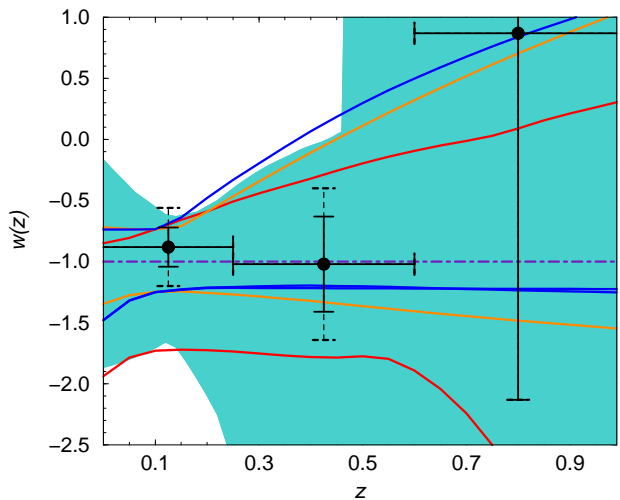}
\caption{The dark energy EOS, $w(z)$, as a function of redshift.
The curves show the 2$\sigma$ allowed values and correspond to the
potentials shown in Fig.~2. Note that if we impose $\d \phi/dz>0$,
then $w > -1$,  as single scalar-field models do not lead to $w <
-1$. However, as shown, most of the parameterizations allow the
region below $w<-1$.  When constructing the potentials shown in
Fig.~2, we apply the condition that $\d \phi/dz>0$. Note that the
$w(z)$ parameterizations, with two free parameters, are the most
restrictive parametrization in the regime $z < 0.3$. Over the
redshift range probed, the different parameterizations generally
agree with each other. The plotted error bars show the 1$\sigma$
and $2\sigma$ errors of $w_i(z)$ when the EOS is subdivided into
three bins in redshift, with $w_i(z)$ directly measured from data
and no restrictions on its values.  A Gaussian prior has been
taken on $\Omega_m$ with a one sigma uncertainty of 0.05  with
$w(z)$ parameterizations.} \label{fig:three}
\end{figure}

\section{Biases in Model-dependent Estimates}

In Fig.~1 we show the Hubble diagram for the 115 data points from
Ref.~\cite{Astier} used in this analysis. For reference, we also
plot $\sim$ 300 distance curves which are 2$\sigma$ consistent
curves drawn from the likelihood distribution for parameters under
the Taylor expansion for $r(z)$.  The best-fit model with this
parameterizations has a chi-square value of 113.1 with 112 degrees
of freedom. Note that in Ref.~\cite{Astier}, $\sigma_{\rm int}$ is
tuned so that $\chi^2=1$ for the best-fit model under
standard-$\Lambda$CDM cosmological fits to the data. We use their
best-fit value, $\sigma_{\rm int}=0.13$, and do not take this
intrinsic uncertainty as an additional free parameter in our
modelling. The exact value of the intrinsic dispersion does not
impact our comparison of different approaches to the
reconstruction of the quintessence potential. It is to be
emphasized that all of our parameterizations of either distance or
the EOS yield comparable $\chi^2$ values for the best-fit model.
This suggests that all four of the reconstruction methods outlined
above are indistinguishable within the redshift range considered.

As discussed in the previous section, for each of the four
parameterizations we determine a best-fit $r(z)$ to the SN data.
We then Monte-Carlo generate models within $2\sigma$ of this
best-fit, generating over 600 instances of $w(z)$ and $V(\phi)$,
all of which are consistent with the underlying SN dataset at the
$2\sigma$ level. In Fig.~2 we show the potentials reconstructed
from each of the four methods, with the bands encapsulating 95\%
of the distribution of the individual models.
Due to the behavior of the Taylor expansion at high $z$, and the
fact that we do not restrict the coefficients of the polynomial
expansion to follow a flat universe, this parametrization gives
rise to a large range of acceptable potentials which satisfy the
data. The Pad\'e parametrization of $r(z)$, as well as the $w(z)$
models, significantly improve the constraints on allowed potential
shapes. This is because the parameters in the Pad\'e approximation
are additionally constrained to satisfy criteria related to the
behavior of $r(z)$ as $z \rightarrow \infty$, as well as by the
assumption of a flat universe \cite{Saini}. When fitting the
$w(z)$ parameterizations to the data, we were able to impose a
prior on $\Omega_m$ based on existing cosmological information
(this was not possible when using $r(z)$ fitting functions). While
we find some overlap in the $2\sigma$ allowed region in the
$\phi$--$V(\phi)$ plane between the four approaches, there are
also noticeable inconsistencies. Analysis of an identical dataset
with different approximations to $V$ or $w$ lead to differing
resulting best-fit potentials.

The differences related to potential shapes between the four
methods are best captured in terms of evolutionary histories for
the dark-energy EOS. In Fig.~3 we summarize the best-fit $w(z)$
results for each of the four reconstruction techniques. Note that
some of our parameterizations allow $w(z) < -1$, but due to our
assumption that the dark energy arises from a scalar-field
potential where $w(z)$ is always expected to be greater than -1,
we restrict the allowed parameter space to be the region where
$w(z) > -1$. Similarly to Fig.~2, we find considerable overlap
between different reconstruction schemes in the $w(z)$ versus
redshift plane, with most models indicating that as the redshift
is decreased, $w(z)$ tends to values between -0.8 and -1.0 at
$z=0$.  In terms of our direct $w(z)$ parameterizations, with
$w(z)=w_0+(1-a)w_a$ we find $w_0 = -1.12 \pm 0.14$ and $w_a=0.38
\pm 0.49$ at the 68\% confidence level.  In the case of
$w(z)=w_0+\alpha \ln (1+z)$ we find $w_0 = -1.08 \pm 0.11$ and
$\alpha=0.35 \pm 0.75$.  As shown in prior studies
\cite{Huterer2}, $w(z)$ parameterizations allow for a minimum
$w(z)$ region at a certain pivot redshift. For the dataset used
here, this pivot redshift is at $z \sim 0.12$, and at the
2$\sigma$ confidence level we find that the pivot point satisfies
$-1.23 < w_p\equiv w(z=0.12) < -0.74$, using $w(z)=w_0+(1-a)w_a$.
It is important to note that all the parameterizations are
consistent with a cosmological constant.

In Figs.~2 and 3 we show the $2\sigma$ bands of best-fit models to
the data, under different parameterizations of the distance or
dark energy EOS. In addition to this outer envelope, we are also
interested in the distribution of the individual $w(z)$ models
within the $2\sigma$ bands. We thus study the behavior of the
models in the $w$--$w'$ plane, which has been suggested as a
natural venue in which to distinguish models~\cite{Caldwell}. We
Monte-Carlo 600 scalar potentials, $V(\phi)$, and evolution
histories, $w(z)$, within the $2\sigma$ regime of the best-fit
parameters for each of the four fitting functions. In Fig.~4 we
plot $w$ and $w'$ at $z=0.1$ and $z=0.5$ for each Monte-Carlo
model, with the scatter of points being $2\sigma$ consistent with
our underlying SN data set.

Based on the evolutionary behavior of simple scalar-field models
in the $w$--$w'$ plane, it has been suggested that one can
separate potentials into ``thawing'' and ``freezing'' regions,
based upon their shapes \cite{Caldwell}. These regions are
delineated in Fig.~4, for comparison with our individual
Monte-Carlo models. It is apparent that the different
parametrization approaches yield separate, though often
overlapping, regions within the $w$--$w'$ plane. In addition, the
models are not necessarily well-contained within the thawing or
freezing regions, with a freezing model in one parametrization
ending up as a thawing model in another, or with models ending up
in between thawing or freezing, or well outside of either regime.
Using generic numerical models for the potential shape, this
behavior has also recently been highlighted in Ref.~\cite{HutPei}.
By applying additional constraints on allowable potentials
(especially at high $z$), Ref.~\cite{Caldwell} find much tighter
confinement in the $w$--$w'$ plane.

Any statement regarding the shape of the scalar potential, as
determined from data, is thus crucially dependent upon the
underlying parameterizations. For example, for the Taylor
expansion approach $w'$ is largely negative at $z=0.5$, while it
is positive at $z=0.1$. Under the Pad\'e approximation, $w'$ is
negative at both $z=0.1$ and $z=0.5$, with $w$ tightly clustered
($-0.9\lap w \lap-0.8$) at $z=0.1$ and relatively unconstrained at
$z=0.5$. Although the reconstructed potentials show significant
overlap (see Fig.~2), the distributions in the $w$--$w'$ plane are
less consistent among different parameterizations. Thus, while
there is motivation from theoretical arguments for using the
$w$--$w'$ plane for potential recognition, there is no obvious,
parametrization independent way to convert distance data to
constraints in this plane.

The differences seen in
Fig.~4 are attributable to the different parametric forms used to approximate
the distance or the dark energy EOS.
To paraphrase our results: you get out what you put in.
Furthermore, the fitting forms to both the distance and the
EOS are motivated by their ability to
fit data, and possess no clear physical motivation.
Fig.~4 thus emphasizes the need for an approach which makes minimal
assumptions about the underlying potential, thereby maximizing the
measurement of a completely unknown scalar field. The less we assume about the
potential, the more powerful the ensuing measurement of its shape.
Such an approach is presented in the following section.

\begin{figure}[tb]
\includegraphics[scale=0.37,angle=-90]{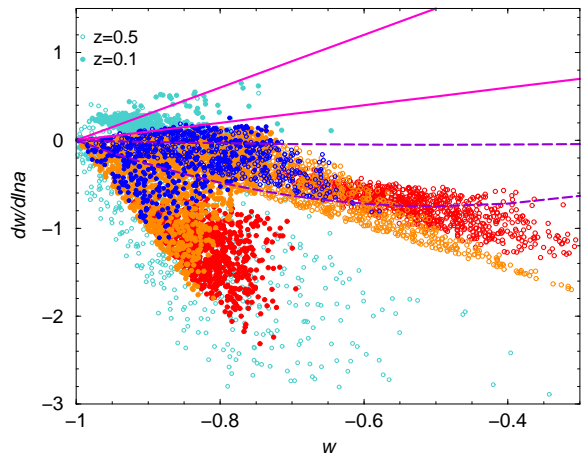}
\caption{$w$ versus $dw/d\ln a$. The data points show the EOS
and its time derivative for 600 model potentials
uniformly drawn at the 2$\sigma$ confidence level at redshift of
0.1 (filled symbols) and 0.5 (open symbols). The cyan, red,
orange, and blue data points show potentials selected under the
Taylor expansion, Pad\'e approximation, $w(z)=w_0+\alpha \ln(1+z)$
and $w(z)=w_0+w_a(1-a)$ model fits to the data, respectively. For
comparison, we also plot the thawing (dashed) and freezing (solid)
potential regions, following Ref.~\cite{Caldwell}. There are
considerable differences in $w$ and $dw/d\ln a$ values, and the
evolution captured by  two redshifts, between the four approaches.
These values do not satisfy the expectations under simple model
criteria for the dark energy potentials, though all of these
potentials, and the $w(z)$ curves, are consistent with the data at
the 2$\sigma$ confidence level.
} \label{fig:four}
\end{figure}

\section{Model-free estimates}

Thus far we have discussed results based on assumed
parameterizations for either distance or dark energy
EOS.  These parameterizations
lead to conclusions that are subject to the assumed
parameterizations.  It is desirable to make model independent
estimates of dark energy. In the case of the EOS
$w(z)$, one could approach this by binning $w(z)$ in redshift
\cite{HutCoo}. Applying this to our SNLS dataset, we evaluate
$w(z)$ over three bins in redshift, $0<z_1<0.25$, $0.25<z_2<0.6$,
and $0.6<z_3<1.0$, assuming $w(z_i)$ constant in each bin.

The resulting best-fit to the SN data is
shown by the data points with $1\sigma$ and $2\sigma$ error bars
in Fig.~3. We find $w(z_1)=-0.88 \pm 0.28$ and
$w(z_2)=-1.02^{+0.94}_{-1.26}$ with no useful constraint for
$w(z)$ in the $z_3$ bin. Although these bins are correlated at the
10\% level, it is possible to decorrelate the binned $w_i(z)$
estimates following the approach of Ref.~\cite{HutCoo}. While only
three bins are attempted here, as SN sample sizes increase, one
can consider larger numbers of bins, each narrower in redshift.
The estimates shown in Fig.~3 are consistent with estimates based
on both fitting functions to the EOS,
 $w(z)=w_0+w_a(1-a)$  and $w(z)=w_0+\alpha \ln (1+z)$.
As discussed in Ref.~\cite{HutCoo}, the binned estimates
capture the dark energy EOS with minimal prior
assumptions on the parameterization. This is expected to maximize
the information one can extract from the data, while minimizing
the introduction of biases.

As discussed and noted elsewhere \cite{Weller,Gerke}, the
scalar-field potential reconstruction is also subject to prior
assumptions on the fitting form. To avoid biases and to make
statements that are not subject to assumed parametrization, it is
useful to directly construct the potential from data.  Recent
approaches in the literature consider fitting distance data to a
potential expanded as a polynomial in the scalar field with
$V(\phi)=\sum_{i=0}^{\infty} V_i \phi^i$
\cite{Sahlen,Sahlen2,HutPei}. Since we are forced to truncate the
expansion at low order (for example, at cubic order with existing
data \cite{Sahlen}), the potential is no longer arbitrary, but
rather has a very limited range of possible shapes.

Instead of assuming a specific family of shapes for $V(\phi)$, we propose
a ``model-free'' extraction of the potential directly from the data.
We make two assumptions about the scalar field potential: (1) that it is a
piecewise continuous function, and (2) that its structure is
``uniform'' in the $\phi$ range explored by the data.
For $(N-1)\Delta\phi<\phi<N\Delta\phi$, we describe the
potential as a
function of the field with constant gradients, $dV/d\phi$,
over binned intervals, $\Delta \phi$:
\begin{equation}
V(\phi) = V_0 + \sum_{i=1}^{N-1} (dV/d\phi)_i \Delta\phi + (\phi - (N-1)\Delta\phi) (dV/d\phi)_N\, .
\label{V_phi}
\end{equation}
Assumption (1) above ensures continuity of $V(\phi)$, which is
necessary since one evolves the potential through the dynamic
equation for the field as $\ddot{\phi} + 3 H
\dot{\phi}+dV/d\phi=0$, and discontinuities would lead to infinite
derivatives. This requirement is unnecessary when considering
parameter-free estimates of the dark energy EOS, which is allowed
discontinuous jumps in redshift. Both the constant-value and the
constant-slope approaches to parameterizing the dark energy EOS
lead to similar conclusions \cite{HutCoo}. Assumption (2) states
that our bins in $\phi$ are fixed width: $\Delta\phi$ is a
constant, independent of $\phi$. This assumption could be relaxed
(e.g., finer bins near $\phi=0$), but this would lead to
additional parameters, in addition to introducing model-dependent
assumptions into the analysis. The expansion of $V(\phi)$ in
Eq.~\ref{V_phi} appears to make the least offensive assumptions
possible, and thereby offers the basis with which to maximally
constrain the full range of possible underlying potentials.

We apply the above potential description to SNLS data following the same
approach as Ref.~\cite{Sahlen}, with three free parameters: $V_0$,
$(dV/d\phi)_1$ for $0<\phi<0.03$, and $(dV/d\phi)_2$ for
$\phi>0.03$. The sizes of the bins are chosen by the range of
$\phi$ we are able to constrain, which is in turn related to
both the redshift range of the SN dataset and the shape of
the potential.
Note that we take
$\phi=0$ to coincide with $z=0$. Instead of $(dV/d\phi)_2$, we
convert the gradient to a data point at $\phi=0.06$, although we
find only an upper limit, as this gradient is not strongly
constrained by existing data.  In Fig.~2 we show the estimated
potential and error bars at the 1$\sigma$ and 2$\sigma$ level. The
potential values allowed by the data are generally consistent with
other indirect reconstructions based on fitting forms for the
distance or the EOS. While fitting forms lead to
largely positive $V(\phi)$ at $\phi>0.05$, our binned approach
finds only an upper limit in this range.

While we have described the potential with only three parameters, this
can be straightforwardly generalized to additional bins as the
statistics and quality of the SN samples improve. In addition, we make a minimal
number of assumptions regarding the potential, and thus are not
biased for or against any particular shapes for the scalar
field potential. The proposed approach
is similar to the case where the EOS is binned and directly measured
from the data without specifying a model for the evolution.  As SN data samples
increase in size, we believe such a model independent approach will become a
powerful tool in extracting information about underlying scalar field
potentials.

\section{Summary}

We have presented a reconstruction of a single scalar-field
potential using recent SN data from the SNLS survey \cite{Astier}.
We have shown that reconstructions based on various approximations
to the distance and the EOS lead to differing
evolution histories of the dark energy EOS, particularly
when the models are examined in the $w$--$w'$ plane.
In this plane the same data can lead to large movements in best-fit models,
depending on the specific approximations to distance or EOS which are being
utilized. Thus the underlying model assumptions lead to biases,
compromising our ability to distinguish evolutionary behaviors of the dark
energy. At present the models are only weakly constrained by the data, and thus
this model-dependence, although apparent, is not a critical failure. As the data
improves, however, a model-independent approach will be essential to determining
an otherwise completely unknown scalar-field potential.

As an alternative to existing indirect reconstruction schemes, we
have thus proposed a technique which establishes the potential
directly from the data, with only minimal assumptions about the
underlying shape of the potential. We take the potential to be a
binned scalar field, piecewise linear and continuous, but
otherwise completely arbitrary. Given the simplicity of these
assumptions, this potential is unlikely to introduce biases in the
determination of a completely unconstrained, underlying potential.
We have demonstrated this approach with current SN data, comparing
the results to parameterized analysis. The ensuing constraints,
although weaker, are expected to be robust and unbiased. It has
been found that direct binning approaches to the dark energy EOS
hold great promise for establishing model-independent measurements
\cite{HutCoo}. We propose a similar approach to reconstructing the
underlying dark energy scalar field potential, allowing us to make
assumption-free statements about the nature of the completely
unknown and mysterious field potentially responsible for the
accelerating expansion of the Universe.

\section{Acknowledgments}
AC and DEH are partially supported by the DOE at LANL
through IGPP grant Astro-1603-07.  DEH acknowledges a
Richard P. Feynman Fellowship from LANL, and is grateful to
the Moore Center for Theoretical Cosmology and Physics at
Caltech for its hospitality. CL is supported by
the Gordon Moore Foundation at Caltech. We thank Dragan
Huterer for useful discussions related to this work, and
Eric Linder and Robert Caldwell for very detailed comments on an
earlier version of the paper.

\end{document}